%% file: a1-2.tex
\documentclass[a4paper]{llncs}

\usepackage[english]{babel}
\usepackage[utf8]{inputenc}
\usepackage{amsmath}
\usepackage{graphicx}
\usepackage[colorinlistoftodos]{todonotes}
\usepackage{tikz}
\usepackage{lscape}
\usepackage{stmaryrd}
\usepackage{eufrak}
\usepackage{mathrsfs}

\numberwithin{equation}{section}
\usepackage{glossaries}
\usepackage{listings}

\loadglsentries{./glossInput.tex}
\makeglossaries
\usepackage{caption}

\usepackage{multicol}
\usepackage[colorinlistoftodos]{todonotes}
\usepackage{cleveref}

\usepackage{xspace} 

\input{macros}

\lstdefinestyle{SSI} {language=C,captionpos=b,tabsize=3,frame=lines,keywordstyle=\color[rgb]{0.6,0,0},morekeywords={or, then, s, End, c, cr, cn, cfr, cfn, xs, l, f, clear, bpull },numbers=left,showtabs=false,morecomment=[l]--,
breaklines=true,stringstyle=\color[rgb]{0.627,0.126,0.941},
basicstyle=\footnotesize\ttfamily}

\lstdefinestyle{SMV} {language=C,captionpos=b,tabsize=2,frame=lines,keywordstyle=\color[rgb]{0.6,0,0},morekeywords={MODULE, VAR, ASSIGN, SPEC, init, next, case,esac,IVAR,self,INVARSPEC,st,rsu},numbers=left,showtabs=false,morecomment=[l]--,commentstyle=\color{gray},
breaklines=true,stringstyle=\color{black},showstringspaces=false,
basicstyle=\footnotesize\ttfamily}
\lstdefinestyle{OTH} {language=C,captionpos=b,tabsize=3,frame=lines,keywordstyle=\color[rgb]{0.6,0,0},morekeywords={SUB,REFINEMENT,COMPONENT,system,
CONNECTION,INPUT,PORT,OUTPUT,INTERFACE,assume,CONTRACT,
guarantee,always,REFINEDBY},numbers=left,showtabs=false,morecomment=[l]--,commentstyle=\color{gray},
breaklines=true,stringstyle=\color[rgb]{0.627,0.126,0.941},
basicstyle=\footnotesize\ttfamily}

\lstdefinestyle{TRA} {language=C,captionpos=b,tabsize=2,frame=lines,keywordstyle=\color[rgb]{0.6,0,0},morekeywords={front,G,status,proving,tisp,st,cmd,L_CS,rsu,Act,rtF,cr,cn,cdr,cdn,Tprogress,t1,t2},numbers=left,showtabs=false,morecomment=[l]--,commentstyle=\color{gray},
breaklines=true,stringstyle=\color[rgb]{0.627,0.126,0.941},
basicstyle=\footnotesize\ttfamily}

\usepackage{float}
\floatstyle{plain}
\newfloat{myequation}{H}{eq}[section]
\floatname{myequation}{Equation}
\usetikzlibrary{arrows,automata,shapes,shadows}
\usetikzlibrary{calc,positioning}

\title{Verification of railway interlocking - Compositional approach with OCRA}
\titlerunning{Verification of railway interlocking}

\author{Christophe Limbr\'{e}e\inst{1} \and Quentin Cappart\inst{1}   \and Charles Pecheur\inst{1} \and\\ Stefano Tonetta\inst{2}}
\institute{Universit\'{e} catholique de Louvain, Louvain-La-Neuve, Belgium
\email{\{quentin.cappart|christophe.limbree|charles.pecheur\}@uclouvain.be}
\and
Fondazione Bruno Kessler, Trento, Italy\\
\email{tonettas@fbk.eu}
}

\date{\today}

\begin{document}

\maketitle

\begin{abstract}
In the railway domain, an electronic interlocking is a computerised
system that controls the railway signalling components (e.g. switches
or signals) in order to allow a safe operation of the train
traffic. Interlockings are controlled by a software logic that relies
on a generic software and a set of application data particular to the
station under control. The verification of the application data is time consuming
and error prone as it is mostly performed by human testers.

In the first stage of our research \cite{my_2}, we built a model of a
small Belgian railway station and we performed the verification of the
application data with the \nusmv model checker. However, the verification of larger stations fails due to the state space explosion problem. The intuition is that large
stations can be split into smaller components that can be verified separately. This concept is known as compositional verification. This article explains how we used the \ocra tool in
order to model a medium size station and how we verified safety properties by mean of  contracts. We also took advantage of new algorithms (k-liveness and ic3) recently implemented in \nuxmv in order to verify LTL properties on our model.
\end{abstract}

\section{Introduction}
\label{intro}
\input{s0_intro}

\section{Contract Based Verification}
\label{formal_techniques}
\input{s2_contract}

\section{System and Model Description}
\label{descModel}
\input{s1_system}

\section{Verification}
\label{verStra}
\input{s3_strategy}

\section{Results and Performance}
\label{results}
\input{s4_case}

\section{Related Work}
\label{relWorks}
\input{s5_related}

\section{Conclusions and Future Work}
\label{conc}
\input{s6_conc}

\renewcommand{\contentsname}{Table of contents}

\addcontentsline{toc}{section}{References}
\bibliographystyle{splncs03}
\bibliography{./reducedBIBBrackets}

\end{document}

%% file: glossInput.tex
\newacronym{SAR}{SAR}{Salle à Relais}

\newacronym{FMEA}{FMEA}{Failure Modes and Effects Analysis}

\newacronym{FMECA}{FMECA}{Failure Modes, Effects and Criticality Analysis}

\newacronym{AMDEC}{AMDEC}{Analyse des modes de défaillance, de leurs effets et de leur criticité}

\newglossaryentry{XSD}
{
  name=XSD,
  description={ (XML Schema Definition) is a World Wide Web Consortium (W3C) recommendation that specifies how to formally describe the elements in an Extensible Markup Language (XML) document}
}

\newacronym{RIOC}{RIOC}{Railway Operation Center}

\newacronym{PLP}{PLP}{Poste à Logique Programmée}

\newacronym{IMC}{IMC}{Interpolation based Model Checking}

\newacronym{ATP}{ATP}{Automatic Train Protection}

\newacronym{ATW}{ATW}{Work announcement}

\newacronym{RAMS}{RAMS}{Reliability Availability Maintenability Safety}

\newacronym{AW}{AW}{Aiguillage Wissel}

\newacronym{LX}{LX}{Level Crossing}

\newacronym{SML}{SML}{SmartLock 400 interlocking system}

\newacronym{CSM}{CSM}{Common Safety Method}

\newacronym{SGM}{SGM}{Serveur Général de Maintenance}

\newglossaryentry{python}
{
  name={Python},
  description={Python is a widely used general-purpose, high-level programming language - \url{https://www.python.org}}
}

\newglossaryentry{LIST}
{
  name={LIST},
  description={In computer science, a list or sequence is an abstract data type that represents an ordered sequence of values}
}

\newglossaryentry{signalman}
{
  name={signalman},
  description={A signalman or signaller is an employee of a railway transport network who operates the points and signals from a signal box in order to control the movement of trains}
}

\newglossaryentry{Scala}
{
  name={Scala},
  description={Programming language for general software applications. Scala has full support for functional programming and a very strong static type system}
}

\newglossaryentry{MySQL}
{
  name={MySQL},
  description={Open-source relational database management system (RDBMS)}
}

\newglossaryentry{MC}
{
  name={MC},
  description={Model Checking}
}

\newglossaryentry{EBP}
{
  name={EBP},
  description={Elektronische Bedieningsposten}
}

\newglossaryentry{TRP}
{
  name={TRP},
  description={Train passage recorder based on a TC and a treadle}
}

\newglossaryentry{IR}
{
  name={IR},
  description={Immobilisation Route}
}

\newglossaryentry{OPT.dat}
{
  name={OPT.dat},
  description={OPT.dat is one of the major file holding the interlocking appliocation data. Especially those of the routes}
}

\newglossaryentry{LX.NG}
{
  name={LX.NG},
  description={Level crossing new generation. Railway application. Infrabel 2014}
}

\newglossaryentry{Prolog}
{
  name={Prolog},
  description={Prolog is a general purpose logic programming language associated with artificial intelligence and computational linguistics. While initially aimed at natural language processing, the language has since then stretched far into other areas like theorem proving}
}

\newacronym{FM} 
{FM} 
 {Formal Methods} 

\newglossaryentry{CTL}
{
  name={CTL},
  description={Computation Tree Logic}
}

\newacronym{OBDDs} 
{OBDDs} 
 {Ordered Binary Decision Diagrams} 

\newacronym{RIS} 
{RIS} 
 {Railway Interlocking System} 

\newglossaryentry{SSI}
{
  name={SSI},
  description={Solid State Interlocking developed by the British Railways}
}

\newglossaryentry{CSP}
{
  name={CSP},
  description={Communicating Sequential Processes (CSP) is a formal language for describing patterns of interaction in concurrent systems.[1] It is a member of the family of mathematical theories of concurrency known as process algebras, or process calculi, based on message passing via channels}
}

\newglossaryentry{BDD}
{
  name={BDD},
  description={In the field of computer science, a binary decision diagram (BDD) is a data structure that is used to represent a Boolean function}
}

\newglossaryentry{UML}
{
  name={UML},
  description={The Unified Modeling Language (UML) is a general-purpose modeling language in the field of software engineering, which is designed to provide a standard way to visualize the design of a system}
}

\newglossaryentry{ADA}
{
  name={ADA},
  description={Ada is a structured, statically typed, imperative, wide-spectrum, and object-oriented high-level computer programming language, extended from Pascal and other languages}
}

\newglossaryentry{DSL}
{
  name={DSL},
  description={A domain-specific language (DSL) is a computer language specialized to a particular application domain}
}

\newglossaryentry{ETCS}
{
  name={ETCS},
  description={The European Train Control System (ETCS) is a signalling, control and train protection system designed to replace the many incompatible safety systems currently used by European railways, especially on high-speed lines}
}

\newglossaryentry{CPN}
{
  name={CPN},
  description={Coloured Petri Net. CPN preserve useful properties of Petri Nets and at the same time extend initial formalism to allow distinction between tokens}
}

\newglossaryentry{INESS}
{
  name={INESS},
  description={INtegrated European Signalling System. The way towards the definition and development of a common core of functional requirements for a new generation of interoperable interlocking systems}
}

\newglossaryentry{ASM}
{
  name={ASM},
  description={In computer science, an abstract state machine is a state machine operating on states which are arbitrary data structures (structure in the sense of mathematical logic)}
}

\newglossaryentry{SNCB}
{
  name={SNCB},
  description={Société Nationale des Chemins de Fer Belges}
}

\newglossaryentry{nusmv}
{
  name={NuSMV},
  description={NuSMV is a symbolic model checker}
}

\newglossaryentry{kripke}
{
  name={Kripke},
  description={A Kripke structure is a variation of non-deterministic automaton proposed by Saul Kripke, used in model checking to represent the behaviour of a system. It is a simple abstract machine (a mathematical object) to capture the idea of a computing machine}
}

\newglossaryentry{CENELEC}
{
  name={CENELEC},
  description={European Committee for Electrotechnical Standardization}
}

\newglossaryentry{Saxby}
{
  name={Saxby},
  description={John Saxby (1821-1913) was a British engineer from Brighton, noted for his work in railway signalling and the invention of the interlocking system of points and signals}
}

\newglossaryentry{PFD}
{
  name={Probability of Failure on Demand},
  description={Mean failure probability in the demand case – the probability that a safety system will not execute its function when it is required to do so}
}

\newglossaryentry{SFF}
{
  name={Safe Failure Fraction},
  description={Proportion of non-dangerous failures – the ratio of the rate of safe faults plus the rate of diagnosed/recognized faults in relation to the total failure rate of the system}
}

\newglossaryentry{ISA}
{
  name=ISA,
  description={Independent safety assessors carry responsibilities that demand a high level of professional conduct. The Codes of Practice of professional bodies whose members may carry out independent safety assessments provide for an appropriate level of general professional conduct}
}

\newglossaryentry{UIC}
{
  name=UIC,
  description={The UIC (French: Union Internationale des Chemins de fer), or International Union of Railways, is an international rail transport industry body (\url{www.UIC.org})}
}

\newglossaryentry{NOBO}
{
  name=NOBO,
  description={Railway Approvals ~ Notified Body (NoBo) and Designated Body (DeBo).Railway Approvals is a Notified Body which assesses compliance to TSI requirements.Following publication of the Railway (Interoperability) Regulations 2011 an organisation called a Designated Body is required to verify compliance to Country Technical Rules}
}

\newglossaryentry{IXL}
{
  name=IXL,
  description={In railway signalling, an interlocking is an arrangement of signal apparatus that prevents conflicting movements through an arrangement of tracks such as junctions or crossings}
}

\newglossaryentry{ATC}
{
  name=ATC,
  description={Automatic Train Control (ATC) is the term for a general class of train protection systems for railways that involves some sort of speed control mechanism in response to external inputs. ATC systems tend to integrate various cab signalling technologies and the use more granular deceleration patterns in lieu of the rigid stops encountered with the older Automatic Train Stop technology}
}

\newglossaryentry{UCL}
{
  name=UCL,
  description={The Université catholique de Louvain (UCL), is Belgium's largest French-speaking university. It is located in Louvain-la-Neuve, with satellite campuses in Brussels, Charleroi, Mons and Tournai}
}

\newglossaryentry{CBTC}
{
  name=CBTC,
  description={Communications-Based Train Control (CBTC) is a railway signalling system that makes use of the telecommunications between the train and track equipment for the traffic management and infrastructure control. This results in a more efficient and safe way to manage the railway traffic}
}

\newglossaryentry{FSM}
{
  name=FSM,
  description={A finite-state machine is a mathematical model of computation used to design computer programs}
}

\newglossaryentry{ONTOLOGY}
{
  name=ONTOLOGY,
  description={Ontology engineering in computer science and information science is a new field, which studies the methods and methodologies for building ontologies: formal representations of a set of concepts within a domain and the relationships between those concepts}
}

\newglossaryentry{RATP}
{
  name=RATP,
  description={The RATP Group (French: Groupe RATP), also known as the Régie Autonome des Transports Parisiens (English: Autonomous Operator of Parisian Transports) is a state-owned public transport operator headquartered in Paris, France}
}

\newglossaryentry{SAT}
{
  name=SAT,
  description={In computer science, satisfiability (often written in all capitals or abbreviated SAT) is the problem of determining if there exists an interpretation that satisfies a given Boolean formula}
}

\newglossaryentry{SATSOLVER}
{
  name=SATSOLVER,
  description={Un solveur SAT est un programme qui décide automatiquement si une formule de logique propositionnelle est satisfaisable}
}

\newglossaryentry{LTS}
{
  name=LTS,
  description={Label Transition System}
}
\newglossaryentry{PLC}
{
  name=PLC,
  description={Programmable Logic Controller}
}
\newglossaryentry{GRIP}
{
  name=GRIP,
  description={Interlockings applications are developed according to processes prescribed by Railway Authorities, such as Network Rail's Governance for Rail Investment Project}
}
\newglossaryentry{LTL}
{
  name=LTL,
  description={In logic, linear temporal logic or linear-time temporal logic (LTL) is a modal temporal logic with modalities referring to time.  In LTL, one can encode formulae about the future of paths, e.g., a condition will eventually be true, a condition will be true until another fact becomes true, etc}
}
\newglossaryentry{Latex}
{
  name=Latex,
  description={LaTeX is a document markup language and document preparation system for the TeX typesetting program. The term LaTeX refers only to the language in which documents are written, not to the editor application used to write those documents}
}
\newglossaryentry{BMC}
{
  name=BMC,
  description={Bounded Model Checking}
}
\newglossaryentry{Ai}
{
  name=Ai,
  description={Ai - Arrondissement - Maintenance area. The Belgian territory is divided in 24 maintenance areas covering the whole Railway Network. Each Ai is in charge of track, signaling and power maintenance}
}
\newglossaryentry{SQL}
{
  name=SQL,
  description={Structured Query Language is a special-purpose programming language designed for managing data held in a relational database management system (RDBMS)}
}

\newglossaryentry{MMI}
{
  name=MMI,
  description={Man Machine Interface - The user interface, in the industrial design field of human–machine interaction, is the space where interaction between humans and machines occurs. The goal of this interaction is effective operation and control of the machine on the user's end}
}
\newglossaryentry{PNC}
{
  name=PNC,
  description={Panne Non Confirmée (french) - Failure not confirmed by the diagnosis at the test bench - Applies to the test of the TFM at Schaerbeek}
}
\newglossaryentry{CLI}
{
  name=CLI,
  description={Centre Logistique Infrastructure (french). Centers or warehouses where the spare parts are located}
}
\newglossaryentry{TCO}
{
  name=TCO,
  description={Total Cost of Ownership. The TCO is the total cost generated by the TFM during its whole life. This takes into account: the purchase, all kinds of maintenance costs and the decommissioning cost}
}
\newglossaryentry{Infrabel}
{
  name=Infrabel,
  description={Infrabel is the Infrastructure Manager of the Belgian Railways}
}
\newglossaryentry{BSRM}
{
  name=BSRM,
  description={Blocage du Sens de Roulement Matérialisé}
}
\newglossaryentry{BSP}
{
  name=BSP,
  description={Blocage du Sens de Roulement Permanent}
}
\newglossaryentry{BSI}
{
  name=BSI,
  description={Blocage du Sens de Roulement Intermittent}
}
\newglossaryentry{CVT}
{
  name=CVT,
  description={Contre-voie}
}

\newglossaryentry{VNS}
{
  name=VNS,
  description={Voie normale}
}

\newglossaryentry{TC}
{
  name=TC,
  description={A track circuit is a simple electrical device used to detect the absence of a train on rail tracks}
}

\newglossaryentry{TFM}
{
  name=TFM,
  description={Trackside Functional Module}
}

\newglossaryentry{MTBF}
{
  name=MTBF,
  description={Mean time between failures (MTBF) is the predicted elapsed time between inherent failures of a system during operation}
}

\newglossaryentry{GMAO}
{
  name=GMAO,
  description={Gestion du Matériel Assistée par Ordinateur - Computerized Aid Management System}
}

\newglossaryentry{RBC}{
 name={Radio Block Centre},
 description={A centralised safety unit working with an interlocking(s) to establish and control train separation. Receives location information via radio from trains and sends movement authorities via radio to trains}
 }
 
\newacronym{CCS} 
{CCS} 
 {Control-Command and Signalling} 
 
\newglossaryentry{software}{
 name={software},
 description={intellectual creation comprising the programs, procedures, rules and any associated documentation pertaining to the operation of a system}
 }
 
\newacronym{SW} 
{SW} 
 {software} 
 
\newglossaryentry{safety}{
 name={safety},
 description={freedom from unacceptable levels of risk}
 }
 
\newglossaryentry{interoperability}{
 name={interoperability},
 description={means the ability of the trans-European high-speed rail system to allow the safe and uninterrupted movement of highspeed trains that accomplish the specified levels of performance}
 }
 
\newglossaryentry{risk}{
 name={risk},
 description={combination of the frequency, or probability, and the consequence of a specified hazardous event}
 }
 
 \newglossaryentry{formal specification}{
 name={formal specification},
 description={a mathematical description of software or hardware that may be used to develop an implementation.}
 }
 
\newacronym{FS} 
{FS} 
 {formal specification} 
 
\newglossaryentry{quality assurance}{
 name={quality assurance},
 description={part of quality management (3.2.8) focused on providing confidence that quality requirements (3.1.2) will be fulfilled}
 }

\newacronym{QA} 
{QA} 
 {quality assurance} 
 
\newglossaryentry{source code}{
 name={source code},
 description={In computer science, source text (or source code) refers to the text of a computer programme written in a programming language that humans can read.}
 }

\newglossaryentry{model-checking}{
 name={model-checking},
 description={model checking explores all possible behaviours of a formal model to determine whether a user specified property is satisfied. In general a model check can have three outcomes. If the model checker determines that all the specified properties are satisfied, it will simply give an affirmative answer. In cases where the property is not satisfied, a counter-example illustrating where and how the property fails to hold is generated automatically. In some cases, a model checker may not be able to determine whether the given property is satisfied or not.}
 }
 
\newglossaryentry{safety regulatory authority}{
 name={safety regulatory authority},
 description={Often a national government body responsible for setting or agreeing the safety requirements for a railway and ensuring that the railway complies with the requirements.}
 }
 
\newacronym{SRA} 
{SRA} 
 {safety regulatory authority} 
 
\newglossaryentry{European Railway Agency}{
 name={European Railway Agency},
 description={The European Railway Agency, the European Community agency for railway safety and interoperability.}
 }
 
\newacronym{ERA} 
{ERA} 
 {European Railway Agency} 
 
\newglossaryentry{hazard}{
 name={hazard},
 description={a condition that could lead to an accident}
 }
 
\newglossaryentry{damage}{
 name={damage},
 description={}
 }
 
\newglossaryentry{company}{
 name={company},
 description={recipient of a maintenance support service provided by the maintenance support service provider}
 }
 
\newglossaryentry{security}{
 name={security},
 description={}
 }
 
\newglossaryentry{danger}{
 name={danger},
 description={}
 }
 
\newacronym{TSI} 
{TSI} 
 {Technical Specification for Interoperability} 
 
\newglossaryentry{European rail traffic management system}{
 name={European rail traffic management system},
 description={Signalling and operation management system using ETCS for the command control and GSM-R for data transmission.}
 }
 
\newacronym{ERTMS} 
{ERTMS} 
 {European rail traffic management system} 
 
\newacronym{SIL} 
{SIL} 
 {safety integrity level} 
 
\newglossaryentry{artifact}{
 name={artifact},
 description={The result of any activity in the software life-cycle such as requirements, architecture model, design specifications, source code and test scripts. A piece of information that is used or produced by a software development process. An artifact can be a model, a description, or software.}
 }
 
\newglossaryentry{formal methods}{
 name={formal methods},
 description={Formal methods are system design techniques that use rigorously specified mathematical models to build software and hardware systems.}
 }
 
\newglossaryentry{conflicting routes}{
 name={conflicting routes},
 description={Interlocked routes that must not be set up at the same time otherwise causing a collision between trains.}
 }

%% file: macros.tex
\newcommand{\ocra}{\textsc{ocra}\xspace}
\newcommand{\nuxmv}{\textsc{nuxmv}\xspace}
\newcommand{\nusmv}{\textsc{nusmv}\xspace}

\newcommand{\opF}{{\ensuremath{\bf F}\;}\xspace}
\newcommand{\opG}{{\ensuremath{\bf G}\;}\xspace}

\newcommand{\durA}[1] {\emph{#1 sec}}

%% file: s0_intro.tex

In the railway domain, an interlocking is a signalling subsystem that controls the routes, the switches and the signals before allowing a train through a station. Computer-based interlockings are configured based on a set of application data particular to each station.
The safety of the train traffic relies on the correctness of the application data.
Usually, the application data are prepared manually and are thus subject to human errors. For example, some prerequisites to the clearance (e.g. green light) of the origin signal of a route can be missing. This kind of error can easily be discovered by a code review or by testing on a simulator. However, errors caused by concurrent actions (e.g. route commands) are much harder to find. In this case, the combination of possible concurrent actions explodes quickly and testing all possible combinations manually is impracticable.
The goal of our research is to develop a method based on model
checking in order to verify the application data. Especially, our
approach must scale-up and allow the verification of real size
interlocking areas.

In a previous work \cite{my_2}, we built a model of a small Belgian
railway station and we performed the verification of the application
data with the \nusmv model checker. However the verification of larger
stations fails due to the state space explosion problem: the models
are too big so that the model checker does not give a result in reasonable
time. In this paper, we therefore tackle the problem with a
compositional approach. The intuition is that large stations can be
split into smaller components that can be verified separately. We
report on the usage of the \ocra tool in order to model a medium size
station and on how we verified safety properties by mean of
contracts. We also took advantage of new algorithms (k-liveness and
ic3) recently implemented in \nuxmv in order to verify LTL
properties on our model.

\paragraph{Outline}

The paper is structured as follows.
In \emph{Section}~\ref{formal_techniques}, we give a brief overview of the
formal techniques that have been used in the case study.
In the \emph{Section} \ref{descModel}, we describe our model and the
new features that we have added compared to our first model. In
\emph{Section} \ref{verStra}, we explain our verification strategy for
larger stations. In \emph{Section} \ref{results}, we discuss the
performance of our verification approach and show how counter examples
are produced when we insert errors in the application data. References
to related work are provided in \emph{Section} \ref{conc}.

%% file: s2_contract.tex


\subsection{Symbolic Model Checking}

Model checking \cite{Clarke1999} is a method to formally verify that a
system is correct. In symbolic model checking~\cite{McM93}, a system
$M$ is described by a finite set $V$ of variables, the initial states
are represented by a formula $I$ over $V$, while the transitions by a
formula $T$ over the variables $V$ and $V'$, where $V'$ represent the
value of $V$ after a transition. In the scope of this paper, we
consider finite-state systems. Thus, without loss of generality, we
can consider $V$ as Boolean variables and formulas in propositional
logic.

A state is an assignment to the variables in $V$. An initial state is
a state that satisfies $I$. A transition is a pair of states that
satisfy $T$. A path is a sequence $\sigma=s_0,s_1,s_2,\ldots$ of
states such that $s_0$ is an initial state ($s_0\models I$) and, for
every $i\geq 0$, $s_i,s_{i+1}$ is a transition ($s_i,s_{i+1}\models
T$). A state $s$ is reachable if there is a path $s_0,s_1,s_2,\ldots$
such that $s=s_i$ for some $i\geq 0$.

In this paper, we specify transition systems in SMV~\cite{McM93}, the
input language of different model checkers such as \nusmv~\cite{nusmv}
and \nuxmv~\cite{nuxmv}. Safety properties have been formalized by
invariants, i.e. formulas over $V$ that must be satisfied by all
reachable states.  Temporal properties have been formalized into
LTL~\cite{Pnu-FOCS77}, which uses temporal operators to specify the
temporal evolution of the transition system. The typical LTL formula
we consider is in the form $\opG(\phi_1\rightarrow F\phi_2)$, where
$\phi_1$ and $\phi_2$ are state formulas over $V$. It means that
whenever $\phi_1$ is true along an execution, $\phi_2$ is true in a
state that follows along the trace.

\subsection{nuXmv: Verification of Components with K-Liveness and IC3}

In the case study we use \nuxmv to prove invariants and LTL
properties. In particular, we use the IC3 algorithm to prove
invariants and the k-liveness algorithm for LTL properties.

IC3~\cite{bradley} is a SAT-based algorithm for the verification of
invariant properties of transition systems.
Very briefly, the idea of IC3 is to build iteratively a sequence of
formulas $F_0,F_1,\ldots,F_k$ such that i) $F_0 = I$, ii) for all $i >
0$, $F_i$ is a set of clauses, iii) $F_i \models F_{i+1}$, iv) $F_i(V)
\land T(V, V') \models F_{i+1}(V')$, and v) for all $i<k$, $F_i
\models P$ where $P$ is the property that we want to verify.
The formulas $F_i$ are therefore over-approximations of the state
space reachable in up to $i$ transitions. They are iteratively
strengthened and extended by generalizing clauses while disproving
candidate counterexamples.
The procedure terminates when either a counterexample is found or when
$F_i=F_{i+1}$ for some $i$ so that $F_i$ is an inductive invariant
that proves $P$.

In \cite{CGMT-TACAS14}, IC3 has been integrated with \emph{predicate
  abstraction} (PA)~\cite{GS97}. The approach leverages \emph{Implicit
  Abstraction} (IA)~\cite{fm09}, which allows to express abstract
transitions without computing explicitly the abstract system, and is
fully incremental with respect to the addition of new predicates.

k-liveness~\cite{kliveness} reduces liveness to a sequence of
invariant checking. It uses a standard approach to reduce LTL
verification for proving that a certain signal $f$ is eventually never
visited ($\opF\opG\neg f$). The key insight of k-liveness is that, for
finite-state systems, this is equivalent to find a $K$ such that $f$
is visited at most $K$ times, which in turn can be reduced to
invariant checking.  k-liveness is therefore a simple loop that
increases $K$ at every iteration and calls a subroutine safe to check
the invariant. In particular, the implementation in \cite{kliveness}
uses IC3 as safe and exploits the incrementality of IC3 to solve the
sequence of invariant problems in an efficient way.

\subsection{OCRA: Contract-Based Compositional Approach}

In this paper, we adopt a compositional contract-based approach and we
use the framework supported by the \ocra tool~\cite{CDT-ASE13}. In
particular, we specify component interfaces in terms of Boolean data
ports and LTL contracts.

The \ocra input language is a component-based description of the system
architecture where every component is associated with one or more
contracts. Each contract consists of an assumption and a guarantee
specified as LTL formulas. The assumption represents a requirement on
the environment of the component. The guarantee represents a
requirements for the component implementation to be satisfied when the
assumption holds.

When a component $S$ is decomposed into subcomponents, the contract
refinement ensures that the guarantee of $S$ is not weakened by the
contracts of the subcomponents while its assumption is not
strengthened. This is checked independently from the actual
implementation of the components and is verified by means of a set of
proof obligations in LTL, which are discharged with model checking
techniques~\cite{CT-SCP15}.

\ocra allows to associate to a component a behavioral model
representing its implementation. The language used for the behavioral
model is SMV. \ocra checks if the SMV model is a correct implementation
of the specified component simply calling NuSMV to verify if the SMV
model satisfies the implication $A\rightarrow G$ for every contract
$\langle A,G\rangle$ of the component.

%% file: s1_system.tex

In this section, we describe the station, the model, and two new features of our model that are the directional locking and the sequential release.

\subsection{The Station}

\begin{figure}[ht]
\centering
\includegraphics[width=1.0\linewidth]{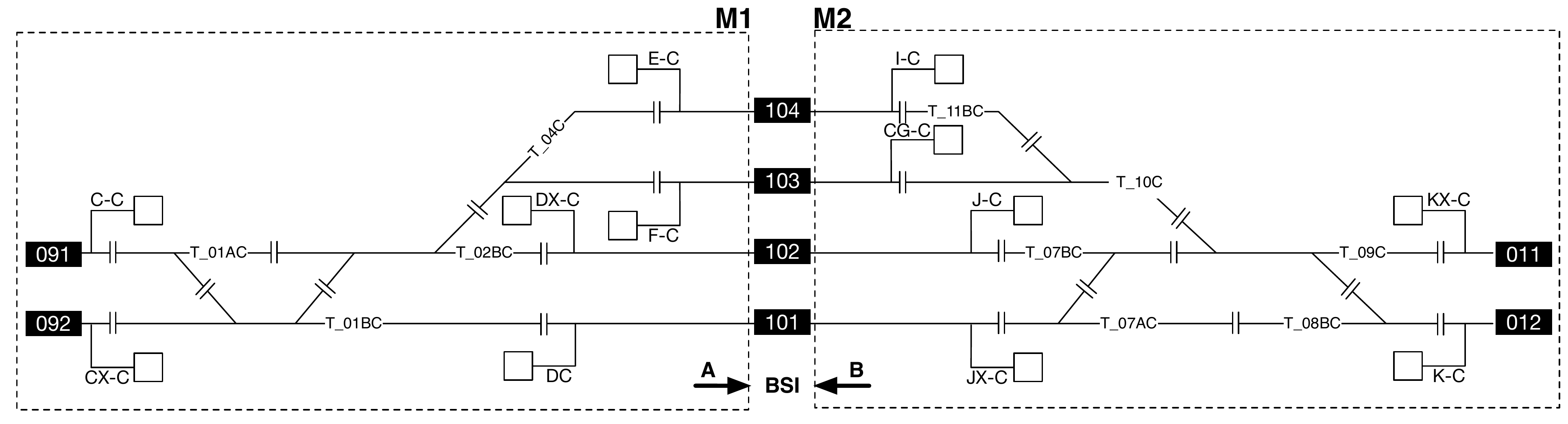}
\captionof{figure}{Track layout of Braine station}
\label{fig:trackLayoutBraine} 
\end{figure}

Braine l'Alleud station, shown in Fig.~\ref{fig:trackLayoutBraine}, is a medium size Belgian railway station comprising $32$ routes, $12$ switches, $12$ signals, and $4$ platforms (101-104). A platform is a section of pathway, alongside rail tracks at a railway station, metro station or tram stop, at which passengers may board. A route is a line of railway track between two signals on a rail system (e.g. route $R\_CC\_102$ from signal CC to track \colorbox{black}{\textcolor{white}{102}} - signal JC). The station can be decomposed into two separate nearly symmetrical parts comprising $16$ routes each: $M1$ and $M2$.

\subsection{Composite System}

The two parts of the station (i.e. $M1$ and $M2$) are not totally
independent but have interfaces. These interfaces materialize a mutual
exclusion mechanism preventing two trains to head for a platform in
opposite direction at the same time (e.g. routes CC\_101 and
KC\_101). Such routes are called conflicting routes. The exercise then
consists in defining the system and its components, the interaction
among the components, and try to prove some global properties on the
system by making assumptions on the environment of each component.

\begin{figure}
\centering
\includegraphics[width=0.99\linewidth]{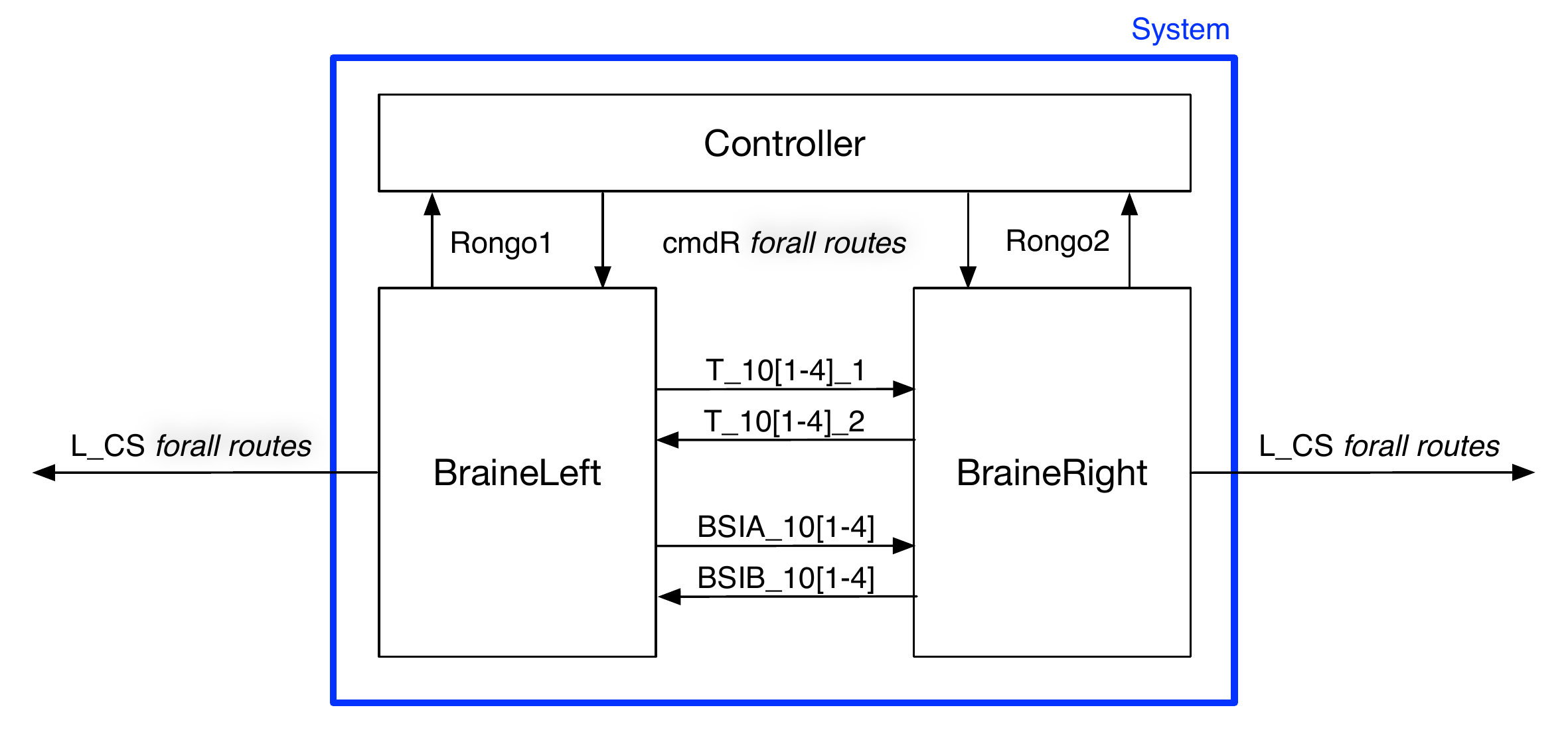}
\captionof{figure}{Architecture of the composite system}
\label{fig:OSS_Braine} 
\end{figure}

The cuts (i.e. $M1$ and $M2$) are chosen so-that: 1) the number of interface variables is minimum, and 2) it sticks to the principle of distribution between interlockings applied in larger stations. The same principle will be applied to two interlockings sharing a section.
As shown in Fig.~\ref{fig:OSS_Braine}, the system is made of three components: $M1$, $M2$, and $C1$. Partial Listing \ref{lst:systRef} shows how the components, and the interfaces are defined in \ocra. The components: $M1$, $M2$, and $C1$ are implemented in SMV language.

\begin{lstlisting}[style=OTH,captionpos=b,caption=System definition in \ocra, label=lst:systRef,frame=none,basicstyle=\small]
COMPONENT BraineLL system

REFINEMENT

  SUB BraineLeft  : M1;
  SUB BraineRight : M2;
  SUB Controller  : C1;

CONNECTION BraineLeft.BSIB_101 := BraineRight.BSIB_101;
...

COMPONENT M1

INTERFACE -- From Environment
  INPUT PORT BSIB_101: boolean;
...

COMPONENT M2

INTERFACE -- From Environment
  OUTPUT PORT BSIB_101: boolean;
...
\end{lstlisting}

The components are defined by means of the $SUB$ keyword. The interfaces are defined as $INPUT$ or $OUTPUT$ (e.g. BSIB\_101 is an output for $M1$ and and an input for $M2$). The $INPUT$ and $OUTPUT$ are connected by mean of the $CONNECTION$ keyword.


Figure~\ref{fig:OSS_Braine} shows how the components are connected by interfaces. The L\_CS OUTPUT variable (=TRUE) is an output of the system and states that the route is set and the origin signal at proceed aspect (e.g. the route $R\_CC\_102$ is set and signal $CC$ is green). The Controller outputs the cmdR variable stating that the controller has issued a route command. The Rongo\{1,2\} INPUT variable provides an acknowledgement that the route command has been properly processed by the interlocking.
The two $M1$ and $M2$ interlocking components exchange the state of the platform track-circuits and the state of the BSI. A track-circuit is an electrical circuit that detects the presence of train in a block of track. The four track-circuits at the platform can be occupied by a train running in either $M1$, or $M2$. The BSI variable allows for mutual exclusion of conflicting routes leading to the same platform. The principle of functioning of the \emph{BSI} is explained in \emph{Sect.} \ref{BSIexplained}.

\subsection{$M1$ and $M2$ models}

\begin{figure}
\centering
\includegraphics[width=0.8\linewidth]{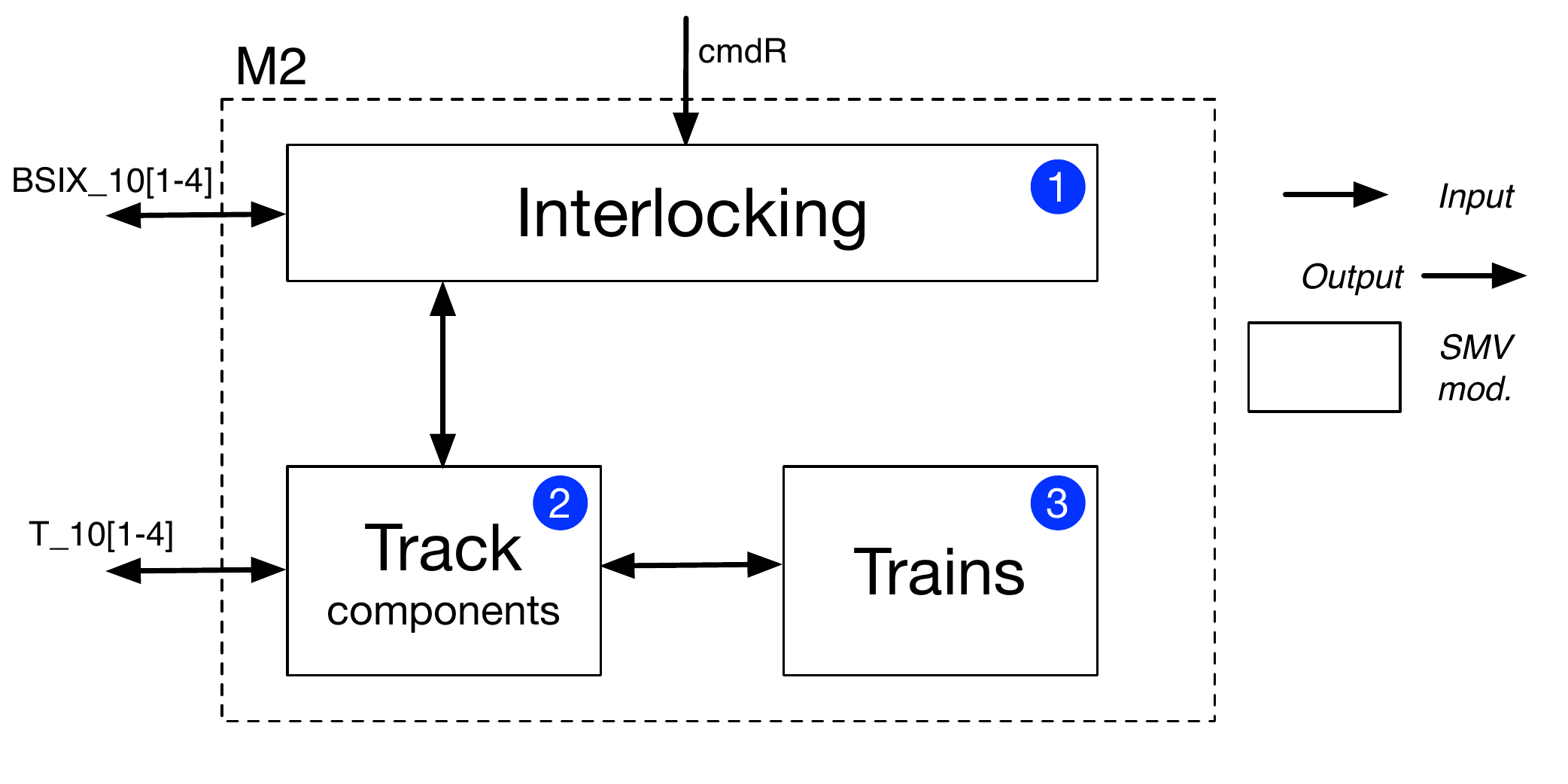}
\captionof{figure}{Architecture of the SMV interlocking model}
\label{fig:modelArch3} 
\end{figure}

Figure~\ref{fig:modelArch3} depicts the internal architecture of the M\{1,2\} component. Each component is implemented in an SMV model. All the modules represent a function achieved by the interlocking except for the train module. In fact the train module allows to simulate the interact of the interlocking with its environment. 



The interlocking module is directly translated from the application data by mean of a translator tool described in \cite{my_2} and models the routes and the locking logic of the switches. Upon a route request, the interlocking (1) verifies that the route can be set and then controls the track components accordingly. A proceed aspect (e.g. green) is sent to the origin signal of the route when the switches are locked in correct position and the track-circuits are clear (i.e. no other train is present on the route). Finally the interlocking detects the trains movement, releases the route and unlocks the resources used by the route. The track components (2) record the status of the track-side objects. For example: for a switch upon a command, the instance verifies that it is not locked before allowing the transition from one position to the other (e.g. left to right). The train modules (3) rely on the track layout of the station. When a signal is at proceed aspect, it simulates a train movement by actuation of the track components. This module is built independently of the application data by mean of a DSL (Domain Specific Language). The train module is local to M\{1,2\} as it is built based on the track layout of its component.

\subsection{BSI Interface Explained}
\label{BSIexplained}

In order to prevent head to head train collisions, the interlocking use a locking mechanism (i.e. $BSI$ - Blocage du sens intermittent in French) that prevent two train to head for the same platform in opposite direction. Fig.~\ref{fig:BSIFSM} illustrates how the $BSI$ variables are actuated upon a route command.

For each platform, two locking variables are used (e.g. BSIA\_102 and BSIB\_102 for platform 2). When no route is set towards platform 102 , the two variables have a permissive value (Free). Upon a route command (e.g. $R\_CC\_102$), the $BSIA\_102$ variable is set in a restrictive state (Locked). The routes in opposite direction (e.g. KC\_102) are thereby blocked and the signal KC can never be commanded to a proceed aspect (e.g. green). The BSIA\_102 variable regains its permissive value when the train has reached platform 102.

\tikzstyle{abstract}=[state, rectangle, draw=black, rounded corners, fill=blue!20, drop shadow, text centered]

\begin{figure}[H]
\centering
\begin{tikzpicture}[->,>=stealth,shorten >=1pt,auto,node distance=3cm,semithick]
  \node (G1) [abstract, rectangle split, rectangle split parts=2]
        {
            BSIA\_102 f
            \nodepart{second} BSIB\_102 f
        };
  \node (G2) [abstract, above right of=G1, rectangle split, rectangle split parts=2]
        {
            BSIA\_102 l
            \nodepart{second} BSIB\_102 f
        };
  \node (G3) [abstract, below right of=G1, rectangle split, rectangle split parts=2]
        {
            BSIA\_102 f
            \nodepart{second} BSIB\_102 l
        };

  \path (G1) edge  [bend left, align=left]            node {R\_CC\_102  cmd} (G2)
    (G1) edge  [bend right, swap]            node {R\_KC\_102  cmd} (G3)
    (G3) edge [bend right, swap, align=left]  node {Train has reached  platform 102} (G1)
    (G2) edge [bend left, align=left]  node {Train has reached  platform 102} (G1);
\end{tikzpicture}

\captionof{figure}{Directional locking for platform 102}
\label{fig:BSIFSM} 
\end{figure}
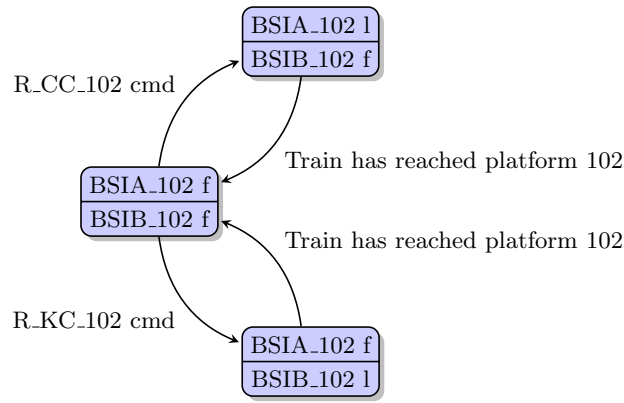

\subsection{Sequential Release}

When the interlocking grants access to a route, it locks all the resources that will be run through by the train: typically all the switches and the track-circuits. This prevents different routes that share those resources to be set at the same time. Such routes are called conflicting routes. Normally those resources are unlocked when the train has completely run through the route. They then become available for other routes. In large stations, it might be interesting to unlock the resources sequentially allowing them to be used by other routes before the train has totally run through. This contributes to improve the train traffic.\\

\begin{figure}
\centering
\includegraphics[width=1.0\linewidth]{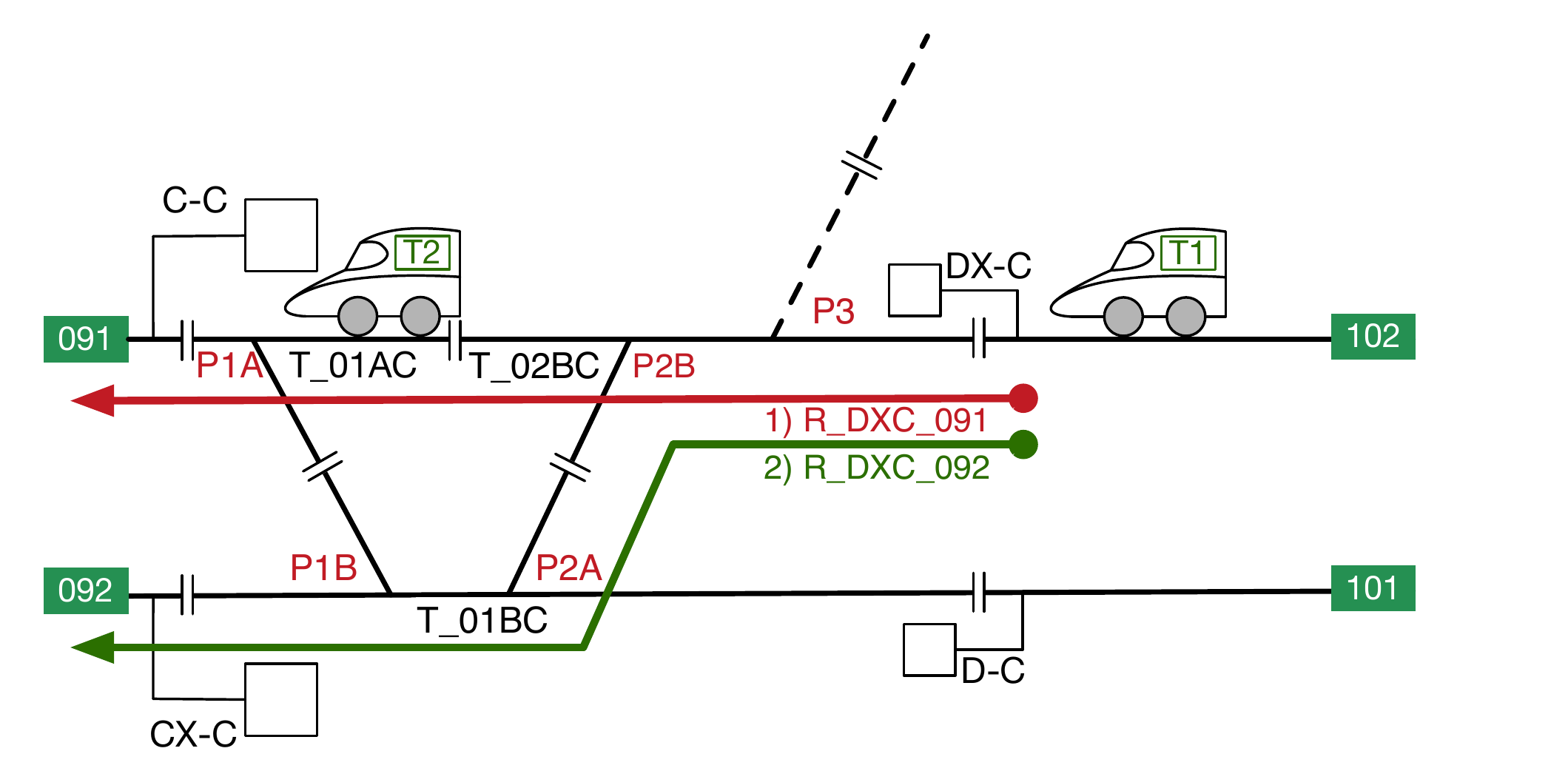}
\captionof{figure}{Sequential release example}
\label{fig:seqRel} 
\end{figure}

The principle of sequential release is illustrated in Fig.~\ref{fig:seqRel}: the first route R\_DXC\_091 is set and prevents the second route R\_DXC\_092 to be set. The following switches are locked: P1A: left, P2B: right, and P3: right. According to the sequential release principle, the second route is set when the train $T_2$ is on the track-circuit T\_01AC and when the track-circuit T\_02BC is free.

%% file: s3_strategy.tex

The decomposition of one interlocking into several components allows
to perform the verification on smaller models (one for each component) and thus limits the so-called state space explosion problem. Therefore we have used two different methods to verify the application data of Braine station: the first takes advantage of the \ocra compositional verification tool and the second uses the \nuxmv tool to verify local properties.
\par The compositional verification applies to the safety properties that imply an interaction between the two components. Those properties are expressed by mean of contracts. An example of contract is given in Listing~\ref{lst:cont1}.
\par A second set of properties are verified straight on each component (i.e. $M1$ and $M2$) with \nuxmv. Several instances of \nuxmv can be started at the same time in order to reduce the computation time of the verification.

\subsection{Compositional Verification}

\subsubsection{Conflicting Routes Controlled by Two Different Components}

The routes R\_CC\_101 and R\_KC\_101 are conflicting because they share the same platform as a destination and the corresponding safety property is expressed by the formula: $P\mbox{ = }G!(R\_CC\_101\_LCS\mbox{ \& }R\_KC\_101\_LCS)$ - $routesTowards\_101$ in Listing~\ref{lst:cont1}. Equation~\eqref{eq:compRule1} shows how this property is verified by composition of the $M1$ and $M2$ modules. The first premise states that when the route R\_CC\_101 is set and origin signal is clear (i.e. R\_CC\_101\_LCS is true), the mutual exclusion property (!BSIA\_101 \& BSIB\_101) is true. The second premise states the same property ($P_2$) for the route R\_KC\_101. $P_1$ and $P_2$ are respectively CC\_101\_OK and KC\_101\_OK in Listing~\ref{lst:cont1}. \ocra performs the verification of these two properties with \nuxmv. The third premise states that the first two premises entail the global property $P$. Finally when all three premises are true, the composition of the components $M1$ and $M2$ verifies the global property $P$.

In other words, if each component (i.e. $M1$ and $M2$) properly blocks the access to a shared platform when it controls a route, then the other component will not be able to control a conflicting route for the same platform.

\begin{myequation}
\[
  \frac{\begin{array}{@{}c@{}}
     \mbox{(Premise 1) }M_1 \models P_1  \\
     \mbox{(Premise 2) }M_2 \models P_2  \\
     \mbox{(Premise 3) }P_1 \wedge P_2 \models P
  \end{array}}{
    M_1\|M_2 \models P}
\]
\caption{Compositional verification of conflicting routes property involving the $M_1$ and $M_2$ components}
\label{eq:compRule1}
\end{myequation}

\begin{lstlisting}[style=OTH,captionpos=b,caption=Contract definition for conflicting routes towards platform 101 involving the $M1$ and $M2$ components,label=lst:cont1,frame=top,frame=bottom,basicstyle=\small]
 CONTRACT routesTowards_101
  assume: always TRUE;
  guarantee: always (R_KC_101_LCS -> !R_CC_101_LCS );

  CONTRACT routesTowards_101
   REFINEDBY M1.CC_101_OK, M2.KC_101_OK;

  CONTRACT CC_101_OK
   assume: TRUE;
   guarantee: always (R_CC_101_LCS -> (!BSIA_101 & BSIB_101));
  CONTRACT KC_101_OK
   assume: TRUE;
   guarantee: always (R_KC_101_LCS -> (!BSIB_101 & BSIA_101));
\end{lstlisting}

\emph{Listing} \ref{lst:cont1} illustrates how the conflicting routes contract for the routes $R\_KC\_101$ and $R\_CC\_101$ is specified. First a top level contract (routesTowards\_101) specifies that the two routes cannot be set at the same time. The top level contract is then refined by two contracts that apply on $M1$ and $M2$: KC\_101\_OK and CC\_101\_OK respectively. These two contracts allow to verify that the $M1$ and $M2$ components handle the $BSI$ locking mechanism properly. The syntax of the language is given in (\cite{OCRA_man}).




\subsection{Local Safety Properties}

The term \emph{Local Properties} designates the properties that are not influenced by the environment of the component on which they are verified.  Those properties are verified on each component SMV model with \nuxmv. Due to the space limitation, those properties will not be explained in detail but examples are provided in Listing~\ref{lst:lProp}. They are expressed in two different ways:
\begin{itemize}
\item By mean of invariants (lines 1 to 5)
\item By mean of $LTL$ formulas and especially by using the ic3 algorithm (lines 6 and 7)
\end{itemize}


\begin{lstlisting}[style=SMV,captionpos=b,caption=Local properties,label=lst:lProp,frame=top,frame=bottom,basicstyle=\small]
check_invar -p "!(M1.t1.front = derailed)"
check_invar -p "!(M1.t1.front = M1.t2.front)"
check_invar -p "!((M1.T_01AC.st = o) & M1.P_01AC.willMove)"
check_invar -p "(M1.R_CXC_103.L_CS ->  !M1.R_EC_091.L_CS)"
check_invar -p "(M1.f1.U_CXC_13C.st = l -> (M1.f1.U_13C_15C.st = l xor M1.f1.U_13C_DXC.st = l))"
check_ltlspec_klive -p "G (M1.U_IR_01AC.st = l -> ((M1.P_01AC.posi = cdr ->  X M1.P_01AC.posi = cdr) & (M1.P_01AC.posi = cdn ->  X M1.P_01AC.posi = cdn)))"
check_ltlspec_klive -p "G((M1.T_01AC.st = o & M1.TRP_CC.krc = s) -> X (!M1.R_CC_101.L_CS & !M1.R_CC_102.L_CS & !M1.R_CC_103.L_CS & !M1.R_CC_104.L_CS))"

\end{lstlisting}

Explanation of the properties:

\begin{itemize}
\item Line 1: the train never derails. A derailment happens when a train takes a trailing point in reverse direction.
\item Line 2: two trains never collide. This is done by verifying that their front never reaches the same track segment at the same time.
\item Line 3: a point never move when its home track-circuit is occupied.
\item Line 4: conflicting routes are not set at the same time. This formula verifies the same property as the contracts defined in \ocra.
\item Line 5: the sub-routes are released in the correct sequence.
\item Line 6: a point never moves when its latching variable is in restrictive state. These formulas are checked by mean of k-liveness (see \cite{klive1})
\item Line 7: signal replacement. The origin signal of a route is immediately commanded to red (replaced) when the train occupies the first track-circuit of the route and has triggered the first passage sensor. This prevents a second train to use the same authorization (i.e. signal green).
\end{itemize}



%% file: s4_case.tex

In this section, we discuss the performance of our verification approach based on composition and local verification. We also illustrate how we validate the model and the properties by error seeding.

\subsection{Performance}

The tests were performed on 2.3 GHz i7 MacBookPro with 4 GB of RAM
running under OS
10.11. Tables~\ref{tab:perf1},~\ref{tab:perf2},~and~\ref{tab:perf3}
illustrate the results (in terms of computation time), which we achieve using different methods and different models. ``BDD'' refers to the fix-point algorithm using BDDs (see \cite{Clarke1999}); ``SAT(ic3)'' refers to the ic3 algorithm using a SAT solver as backend (see \cite{bradley}); ``SMT(ic3)'' refers to the
ic3 algorithm integrated predicate abstraction using an SMT solver as back-end (see \cite{CGMT-TACAS14}).


\begin{table}[H]
\centering
\caption{Performance of the verification of invariants on monolithic models}
\begin{tabular}{|c|c|l|c|c|c|}
\hline
\multicolumn{2}{|p{4.5cm}|}{Model}        & Tool        & Method     & Properties                      & Duration \\
\hline
1  & Monolithic model                  & NuSMV    & BDD      & Invariants                      & $> 1 \mbox{ }day$ \\ \hline
2  & Partial monolithic model         & NuSMV    & BDD       & Invariants                      & $> 1 \mbox{ }day$ \\ \hline
3  & Monolithic model                 & NuXMV    & SAT(ic3)  & 10 $\times$ Invariants      &  \durA{123} \\ \hline
4  & Monolithic model                 & NuXMV    & SMT(ic3)  & 10 $\times$ Invariants      &  \durA{80} \\ \hline
\end{tabular}
\label{tab:perf1}%
\end{table}

Table~\ref{tab:perf1} reports the performance of the verification of the application data for the station described in Section~\ref{descModel}. Line 1 shows that \nusmv could not terminate in one day. After reducing the size the state space by allowing only $16$ routes to be commanded, \nusmv could build the reachable state space in $6$ days and verify invariants (line 2). One of the features of \ocra is to allow to rebuild a monolithic ($32$ routes) model based on the definition of the system. The verification of invariants is therefore possible.
The results clearly show that ic3 with predicate abstraction performs
better than plain ic3, and that both outperform the BDD-based
algorithm on this case study.

\begin{table}[H]
\centering
\caption{Performance of the verification of the contracts by \ocra}
\begin{tabular}{|c|c|l|c|c|c|}
\hline
\multicolumn{2}{|p{4.5cm}|}{Model}        & Tool        & Method     & Properties                      & Duration \\
\hline
5  & Contract refinement              & \ocra        & -           & 4 $\times$ Contracts        &  \durA{7,34} \\ \hline
6  & Implementation M1              & \ocra        & ic3         & 4 $\times$ Contracts        &  \durA{5,6} \\ \hline
7  & Implementation M2              & \ocra       & ic3         & 4 $\times$ Contracts         &  \durA{14,94} \\ \hline
8  & Composite monolithic           & \ocra        & ic3         & 4 $\times$ Contracts          & \durA{1242} \\ \hline
\end{tabular}
\label{tab:perf2}%
\end{table}

Table~\ref{tab:perf2} illustrates the performance of the verification of the contracts and their implementation. Line 5 corresponds to the verification of the premise 3 of \emph{Equation} \eqref{eq:compRule1} ($P_1 \wedge P_2 \models P$). Lines 6 and 7 are respectively related to the verification of the premisses 1 and 2 ($M_1 \models P_1$ and $M2 \models P_2$). The sum of the duration of these 3 tasks gives the time needed by \ocra to check the coherence between the contracts and their implementation in the SMV models (i.e. $\leq$ \durA{28}). This time is to be compared with the \durA{1242} needed by \ocra to verify the same contracts and implementations on a monolithic model.

\begin{table}[H]
\centering
\caption{Performance of the verification of the local properties}
\begin{tabular}{|c|c|l|c|c|c|}
\hline
\multicolumn{2}{|p{4.5cm}|}{Model}        & Tool        & Method     & Properties                      & Duration \\
\hline
9  & M1                               & NuXMV    & BDD       & 197 $\times$ Invariants      &  \durA{123} \\ \hline
10 & M2                               & NuXMV   & BDD        & 199 $\times$ Invariants      &  \durA{424} \\ \hline
11 & M1                              & NuXMV    & SAT(ic3)  & 12 $\times$ LTL               &  \durA{960}\\ \hline
12 & M1                              & NuXMV   & SMT(ic3)   & 12 $\times$ LTL             &  \durA{20} \\ \hline
13 & M2                              & NuXMV   & SAT(ic3)   & 12 $\times$ LTL             & \durA{1036} \\ \hline
14 & M2                              & NuXMV   & SMT(ic3)  & 12 $\times$ LTL             &  \durA{740} \\ \hline
\end{tabular}
\label{tab:perf3}%
\end{table}

Finally Table~\ref{tab:perf3} illustrates the verification of the
local safety properties on the $M1$ and $M2$ components. Two
approaches are used: first some invariants are verified with \nuxmv
and the standard BDD and second 12 LTL properties are verified with
the ic3 algorithm. An order file based on \cite{WIN1} is used to
optimize the BDD structure. ic3 with abstraction and the SMT MathSAT
solver outperforms ic3 with MiniSAT in this context in an order of
magnitude close to $50$.

\subsection{Error Seeding}

In order to gain confidence in our model and properties, we have seeded errors in the model by removing some safety conditions in the route proving conditions\footnote{Conditions to give a proceed aspect on origin signal of the route.}. As expected, \ocra could not prove the safety property and produced a counterexample. Listing~\ref{lst:trace01} shows that the property is false (line 1) because the route R\_KXC\_101 is set (line 30) whereas the BSIB\_101 is TRUE (line 5). 

\begin{multicols}{2}
\begin{lstlisting}[style=TRA, caption=Error trace generated after error seeding in the model, label=lst:trace01,frame=none,basicstyle=\small]
LTL spec G (R_KXC_101_LCS -> (!BSIB_101 & BSIA_101)) is false
Trace Description: IC3 counterexample
  -> State: 2.1 <-
...
    BSIB_101.st = TRUE
...
  -> Input: 2.2 <-
    cmdR = R_KXC_101
  -> State: 2.2 <-
    R_KXC_101.cmd = TRUE
...
  -> Input: 2.3 <-
    cmdR = R_KXC_103
  -> State: 2.3 <-
    R_KXC_101.st = s
...
  -> State: 2.4 <-
    U_IR_09C.st = l
    U_IR_07BC.st = l
    BSIA_101 = TRUE
    U_IR_07AC.st = l
    U_16C_JXC.st = l
    U_18C_16C.st = l
    U_KXC_18C.st = l
...
  -> State: 2.5 <-
    R_KXC_101.st = rsu
    T_101.st = c
    T_101_1 = FALSE
    R_KXC_101_LCS = TRUE (Route is set)
    KXCopen = TRUE

\end{lstlisting}
\end{multicols}

%% file: s5_related.tex

Many works applied model checking to interlocking systems. One of
the first work dates back to 1998 and is described
in \cite{Spin1}. However, as also concluded in \cite{FMGF-FORMAT10},
although small scale interlocking systems can be addressed by model
checking, interlockings that control medium or large railway needs to
tackle the state-space explosion problem. As shown also
in \cite{CCLNRRST-CAV12}, a single approach is often not sufficient to
prove all properties and sometimes a combination of approach may
dramatically improve the performance.

Compositional approach is one method to reduce the complexity of the
verification but is not the only one. For instance, Cappart et al. \cite{cappart_simu} introduced a method based on discrete event simulations. The idea is to do not verify all the states but to limit the verification to a set of likely scenarios. However  this method does not provide enough confidence that all the errors in the application data will be detected.

In \cite{Winter1_1,Winter1}, Winter shows how to compute optimized variable and transition orderings in order to speed-up the symbolic model checking of railway interlockings with NuSMV. She also reported on her findings on how to set the threshold for cluster.

In \cite{WIN1}, Winter et al. modelled the interlocking by means of the formal notation ASM that are more readable. The formal model is translated in NuSMV code and the Safety requirements are expressed in CTL.

In \cite{IRSEProver}, Peter Duggan (Siemens Rail Automation, UK) and Arne Bor{\"a}lv (Prover Technology AB, Sweden) have demonstrated that the Prover\footnote{\url{http://www.prover.com}} tools were successfully used to generate and test the configuration data of a realistic size UK station.

In \cite{LUX1}, Haxthausen et al. detailed how they modelled an ETCS level 2 compatible Danish interlocking with the RT-Tester. The state space, the transition relation and the safety properties are efficiently evaluated by the SMT solvers that support bit vector and integer arithmetic. The model also include the sequential release feature.

In \cite{RBCcompVerif}, Xu et al. verifies hybrid safety properties of Automatic Collision Avoidance System (ACAS) in the European Train Control System (ETCS). They verify those properties using Compositional Verification rules based on weakly monotonic time extension.

In \cite{SNCFpetri}, Antoni et al. have developed a SIL4 interlocking that uses the Petri Nets as application data. In \cite{rssrPetri}, Dutilleul et al. have also used the Petri Nets in order to define a model pattern of railway interlocking.

%% file: s6_conc.tex

\subsection*{Conclusions}

The verification of medium and large interlocking data is still a
challenge due to the state space explosion problem affecting the model
checking process. Our main contribution was to achieve the verification
of the application data of a medium size railway interlocking by mean
of compositional verification. In order to do that, we modelled our
case study interlocking as a composite of smaller interlocking
components in \ocra and SMV language. The verification of the safety
properties (expressed as contracts) was performed with \ocra and
\nuxmv tools.

We have also added the sequential release functionality into our interlocking model. This functionality allows to increase the throughput of the railway network by releasing the route components earlier.

Finally, we have achieved the verification of LTL properties in efficient time thanks to the usage of the new ic3 algorithm implemented into \nuxmv. The verification of the local components can be paralleled by running several instances of \nuxmv at the same time.

\subsection*{Future work}

In our future work, we will continue to refine the structure of the interlocking composite into adequate components (e.g. train). Our goal is to be able to verify safety properties on a network of interlockings by mean of compositional verification.

We will continue to develop the automatic translator tool in order to convert the application data of a network of interlockings into \ocra language.

Another goal is to develop a model of an IL/ETCS installation in order to verify safety properties related to the train dynamic characteristics (i.e. speed and position). In order to do that we will extend our train module in order to make it continuous.